\documentclass[12pt]{article}
\usepackage{amsmath}
\begin{document}
\centerline {\Large{\bf A Closed Universe with Maximum Life-Time}}
\centerline{}
\centerline{\bf {Salah Haggag $^{(1,2)}$, Samy A. Abdel-Hafeez $^{(3)}$, Moutaz Ramadan $^{(3,4)}$}}
\centerline{$^{(1)}$ Department of Basic Sciences, Egyptian Russian University}
\centerline{$^{(2)}$ salah-haggag@eru.edu.eg}
\centerline{$^{(3)}$ Department of Mathematics and Computer Science,}
\centerline{ Faculty of Science, Port Said University, Egypt}
\centerline{$^{(4)}$ motaz\_ramadan@sci.psu.edu.eg}
\centerline{}
\begin{abstract} 
\noindent
A closed universe with maximum life-time is constructed using optimal control. Einstein's field equations are used with varying cosmological "constant". The second time derivative of the Hubble parameter acts as the control function in the optimal control model.
\end{abstract}
\section{Introduction}
In standard cosmology, a spatially homogeneous and isotropic universe is described by the Friedmann-Robertson-Walker (FRW) line element
\begin{equation}\label{m6eq1}
ds^{2} = dt^{2}-a^{2}(t)\left[\frac {dr^{2}}{1-k r^{2}}+r^{2} \left(d\theta ^{2}+\sin^{2} \theta d \phi^{2}\right)\right], ~
\end{equation} 
where $k =-1,0,1$ corresponds to open, flat and closed universe respectively, $r , \theta$ and $\phi$ are spherical polar coordinates, $t$ is the cosmic time, and $a(t)$ is the scale factor.\\
We consider Einstein's field equations with a varying cosmological "constant"  
\begin{equation}\label{m6eq2}
R_{ij}-\frac{1}{2}g_{ij} R= 8\pi G T_{ij}+\Lambda(t) g_{ij}, 
\end{equation}
where $R_{ij}$ is the Ricci tensor, $T_{ij}$ is the energy-momentum tensor that describes the matter distribution, and $\Lambda$ is a cosmological constant . For a perfect fluid with isotropic pressure $p$ and energy density $\rho$, $T_{ij}$ has the form 
\begin{equation}\label{m6eq4}
T_{ij} =-pg_{ij}+(p+\rho )U_{i}U_{k}~,
\end{equation}
where $U_{i}$ is the four-velocity vector with $U_{i}U^{i}=1$.\\
For the FRW metric (\ref{m6eq1}), Einstein's field equations (\ref{m6eq2}) reduce to two equations
\begin{equation}\label{m6eq5}
3H^2+{\frac{3k}{a^2}}= 8\pi G \rho+\Lambda,       \hspace{1cm}
\end{equation} 
\begin{equation}\label{m6eq6}
2 \dot {H}+3H^2+{\frac{k}{a^2}}=-  8\pi G \alpha \rho+\Lambda,   \hspace{1.5cm}
\end{equation} 
where an overdot denotes the derivative with respect to $t$, and $H(t)$ is the Hubble parameter defined by
\begin{equation}\label{m6eq7}
H:=\frac {\dot a}{a}. \hspace{1cm}
\end{equation} 
In standard cosmology, $p$ and $\rho$ are usually assumed to be related by the baratropic equation of state 
\begin{equation}\label{m6eq8}
p=\alpha \rho,     \hspace{2cm}   \alpha={\text {constant}}  \;\;,\;\;   0\leq \alpha \leq 1.~
\end{equation} 
From Eqs.(\ref{m6eq5}),(\ref{m6eq6}), we obtain
\begin{equation}\label{m6eq10}
2\ddot {H}+6H \dot H (1+\alpha )-{\frac{2k}{a^2}}H (1+3\alpha)=\dot \Lambda (1+\alpha). \hspace{1cm}
\end{equation} 
We may take the field equations as the two independent equations (\ref{m6eq7}) and (\ref{m6eq10}) in three unknowns $H(t),~a(t)$ and $\Lambda (t)$.\\
In order to obtain solutions, authors assumed various forms of $\Lambda$. Ozer and Taha~\cite{Ozer86,Ozer87} have suggested a cosmological model on the assumption that the energy density is always at its critical value $\rho_c$ which yields explicit dependence of $\Lambda$ on a scale factor as $\Lambda=\alpha/a^2,~ \alpha=3/8\pi G$ for $k=+1$. Chen and Wu~\cite{Chen} have proposed the cosmological models is proportional to scale factor as $\Lambda \propto a^{-2}$. Berman \cite{Berman91} has proposed $ \Lambda \propto t^{-2}$. Al-Rawaf and Taha~\cite{RawafT}, Al-Rawaf \cite{Rawaf} and Overduin and Cooperstock \cite{Overduin} have assumed that $\Lambda \propto ({\ddot a}/a)$. Arbab~\cite{Arbab2003a,Arbab2003b} has investigated the cosmological implications of a decay law for $\Lambda $ that is proportional to $({\ddot a}/a)$ or $H^2$ or $\rho$.\\
In this paper, rather than using an arbitrary assumption for $\Lambda$, we use optimal control \cite{Pontryagin} to find a new solution for a closed universe, where $\Lambda(t)$ is determined by some interesting criterion. It is well known that a closed universe, with $k=+1$, expands from zero to a maximum size, then contracts back to zero. A closed universe has a finite life-time. It is interesting to {\it{design}} a closed universe with maximum life-time.\\
An optimal control problem is to determine the evolution of a dynamical system such that a specific objective function is minimized or maximized. Optimal control has many applications in diverse areas such as engineering, robotics, finance, economics, biology and other areas. However, it has quite a few applications in astrophysics and cosmology. Rhoades and Ruffini~\cite{Rhoades} and Kalogera and Baym~\cite{Kalogera} used optimal control to establish a bound on the maximum mass of a neutron star with different sets of assumptions. Haggag and Safko~\cite{Haggag} used optimal control to obtain a concise derivation of the Tolman-Oppenheimer-Volkoff equation of hydrostatic equilibrium. Pope~\cite{Pope} investigated an optimal control model of AGN feedback in massive galaxies and galaxy clusters. Haggag {\it {et al.}} \cite{Haggag1} obtained a slow-roll inflationary model.
 
In Section 2, an optimal control problem is constructed for a closed universe with maximum life-time. In Section 3, we derive the solution of the problem using Pontryagin's maximum principle. A brief outline of such an approach is given in \cite{Haggag}. 
\section{The Optimal Control Problem}
When $k=+1$ equations (\ref{m6eq7}) and (\ref{m6eq10}) reduce to
\begin{equation}\label{m6eq3.1}
\dot {a}=aH, \hspace{1.8cm}
\end{equation} 
 \begin{equation}\label{m6eq3.2}
\dot \Lambda=\frac{2}{(1+\alpha)}\ddot {H}+6H \dot H -{\frac{2}{a^2}}\left(\frac{ 1+3\alpha}{1+\alpha}\right)H, \hspace{1cm}
\end{equation}
which are two independent equations in three unknowns  $H(t),a(t)$ and $\Lambda(t)$.\\ 
Further, we need to specify boundary conditions. In order to avoid the Big Bang singularity, we take the initial time, $t=t_i=0$, shortly after the Big Bang moment, when the size of the universe becomes little greater than zero, namely $a(0)=a_0>0$. If we also assume $H(0)=0$, then expansion requires $\dot{H}(0)=A>0$, $A$ is constant. On the other hand, we take the terminal time $t=t_f=T$ when expansion stops, and the size of the universe reaches its maximum, namely $H(T)=0$ and $\dot{H}(T)<0$.\\
To construct an optimal control model, let
\begin{equation}\label{m6eq3.1_1}
w:=\dot H, \hspace{1.8cm} {\text{ and}} \hspace{1.8cm} u:=\dot w =\ddot H.
\end{equation} 
We take $\Lambda(t), H(t), a(t)$ and $w(t)$ as the state functions, and $u(t)$ as the control function. Then, with $T$ as half the life-time, the optimal control problem is to determine the control function $u(t)$ which maximizes the objective function 
\begin{equation}\label{m6eq3.3}
~~J ={\int \limits_{0}^{T} ~dt }, \hspace{1cm} 
\end{equation} 
subject to
\begin{eqnarray}\label{m6eq3.3_1}
\dot \Lambda &=&\frac{2}{(1+\alpha)}u+6H w -{\frac{2}{a^2}}\left(\frac{ 1+3\alpha}{1+\alpha}\right)H, \hspace{0.5cm} 
\end{eqnarray} 
\begin{eqnarray}\label{m6eq3.3_2}
\dot a&=a H, \hspace{3.1cm} a(0)=a_0, \hspace{0.8cm}
\end{eqnarray}
\begin{eqnarray}\label{m6eq3.4}
\dot H&=&w, \hspace{2.9cm}    H(0)=H(T)=0,
\end{eqnarray} 
\begin{eqnarray}\label{m6eq3.5}
~~~~~~~~~~~~~~~\dot w& =& u, \hspace{3.2cm}   w(0)=A>0,~~~~~~~~~ |u|\leq 1.
\end{eqnarray} 
The Hamiltonian for the above system is given by 
\begin{eqnarray}\label{m6eq3.6}
S(t) =1+\lambda_1 \left[\frac{2}{(1+\alpha)}u+6H w -{\frac{2}{a^2}}\left(\frac{ 1+3\alpha}{1+\alpha}\right)H\right]+\lambda_2 a H+\lambda_3 w+\lambda_4  u.   
\end{eqnarray} 
The adjoint equations are
\begin{eqnarray}\label{m6eq3.7}
\dot \lambda_1&=&-{\frac{\partial S}{\partial \Lambda}} =0, \hspace{3.5cm} \lambda_1(T) =0,
\end{eqnarray} 
\begin{eqnarray}\label{m6eq3.8}
~\dot \lambda_2 &=&-{\frac{\partial S}{\partial a}}=-{\frac{4}{a^3}}\left(\frac{ 1+3\alpha}{1+\alpha}\right)H \lambda_1-H \lambda_2, \hspace{.8cm} ~
\end{eqnarray} 
\begin{eqnarray}\label{m6eq3.8_1}
~\dot \lambda_3 &=&-{\frac{\partial S}{\partial H}}=\left[{\frac{2}{a^2}}\left(\frac{ 1+3\alpha}{1+\alpha}\right)-6w\right] \lambda_1-a \lambda_2, \hspace{0cm} ~
\end{eqnarray} 
\begin{eqnarray}\label{m6eq3.8_2}
\dot \lambda_4 &=&-{\frac{\partial S}{\partial w}}=-6H \lambda_1- \lambda_3, \hspace{3.2cm} ~
\end{eqnarray} 
where $\lambda_1$,  $\lambda_2$,  $\lambda_3$ and $\lambda_4$ are costate functions associated with the states $\Lambda$, $a$, $H$ and $w$ respectively.
Since the final time is free, we have $S(T)=0$ and hence
\begin{eqnarray}\label{m6eq3.8_30}
 \lambda_4(T) &=&-\frac{w(T)\lambda_3(T)+1}{u(T)}. \hspace{3.2cm} ~
\end{eqnarray}
\section{A Closed Universe with Maximum Life-Time}
We note that $S(t)$ is linear in $u(t)$ with the factor
\[m=\lambda_4+\frac{2}{(1+\alpha)}\lambda_1.\]
Thus, the Hamiltonian $S$ is maximized with respect to u by taking
\[
u(t) =
  \begin{cases}
   -1       & \quad \text{if } m <0\\
   +1  & \quad \text{if } m>0\\
  \end{cases}
\]
If at some instant of time, $m$ changes sign, then $u$ switches to the other boundary. This is called bang-bang control \cite{Naidu}. The number of switches between boundaries will depend on the number of sign changes in $m$.\\
From Eq.(\ref{m6eq3.7}), we obtain
 \begin{equation}\label{m6eq3.8_3}
 \lambda_1 =C_1, \hspace{4.cm} C_1={\text{Const.}}
\end{equation}
 Taking the time derivative of Eq.(\ref{m6eq3.8_2}) and using Eqs.(\ref{m6eq3.4}), (\ref{m6eq3.7}) and (\ref{m6eq3.8_1}), we obtain
\begin{equation}\label{m6eq3.9}
{\ddot \lambda_4} ={{-2C_1}\frac{-2C_1}{a^2}}\left(\frac{ 1+3\alpha}{1+\alpha}\right) +a\lambda_2. \hspace{2.3cm}
\end{equation}
Differentiating Eq.(\ref{m6eq3.9}) with respect to $t$ and using Eqs.(\ref{m6eq3.3_2}) and (\ref{m6eq3.8}), we obtain
\begin{equation}\label{m6eq3.10}
{ \lambda_4^{(3)}}(t) =0.~
\end{equation}
Therefore,
\begin{equation}\label{m6eq3.11}
~~~~~~~~\lambda_4(t) =\frac{1}{2}C_2~t^2 +C_3 t+C_4,   
\end{equation}
where $C_2,C_3,C_4$ are arbitrary constants. Comparing the time derivative of Eq.(\ref{m6eq3.11}) with Eq.(\ref{m6eq3.8_2}), we obtain
\begin{equation}\label{m6eq3.12}
\lambda_3=-6C_1H-C_2t-C_3.  \hspace{.3cm}
\end{equation}
 Comparing the time derivative of Eq.(\ref{m6eq3.12}) with Eq.(\ref{m6eq3.8_1}), we obtain
\begin{equation}\label{m6eq3.12_1}
\lambda_2=\frac{2C_1}{a^3}\left(\frac{ 1+3\alpha}{1+\alpha}\right)+\frac{C_2}{a}
\end{equation}
From Eqs.(\ref{m6eq3.8_3}) and (\ref{m6eq3.11}), we obtain
\begin{equation}\label{m6eq3.13_0}
m=\frac{2C_1}{(1+\alpha)}+\frac{1}{2}C_2~t^2 +C_3 t+C_4,  \hspace{0.1cm}
\end{equation}
which has at most two sign changes depending on the constants $C_n,~n=1,\dots,4$. 
Now, substituting $u = {\pm 1}$, the system state equations (\ref{m6eq3.4}) and (\ref{m6eq3.5}) reduce to 
\begin{eqnarray}\label{m6eq3.13}
\dot w&=&\pm 1,   \hspace{2.cm}
\end{eqnarray} 
\begin{eqnarray}\label{m6eq3.14}
\dot H =w=\pm t+A.  \hspace{1.2cm}
\end{eqnarray}
Integration of equation (\ref {m6eq3.14}) gives Hubble's parameter
\begin{eqnarray}\label{m6eq3.15}
H(t)& =& \pm\frac{1}{2} t^2+At,
\end{eqnarray}
with the initial condition $H(0)=0$.
Since the expansion halts at $t=T$, then $H(T)=0$ and we obtain 
\begin{equation}\label{m6eq3.23}
T =\mp2A, ~~~~~~~~~~{\text{for}}~~~u=\pm 1.
\end{equation}
Since $A>0$ and $T>0$, the optimal solution is given by
\begin{subequations}\label{m6eq3.35}
\begin{align}
 u&=-1, \\
 H &=-\frac{1}{2} t^2+At, \\
a(t)&= a_0 \exp \left[ \frac{-1}{6} t^2(t-3A)\right].
\end{align}
\end{subequations}
The maximum life-time is
\begin{equation}\label{m6eq3.26}
2T=4A.
\end{equation}
The maximum value of the scale factor is
\begin{equation}\label{m6eq3.28}
a_{max}=a(T)= a_0 \exp\left[\frac{ 2}{3} A^3\right].
\end{equation}
Using Eqs.(\ref{m6eq5})-(\ref{m6eq8}), we obtain
\begin{equation}\label{m6eq10_2}
\Lambda=\frac{1}{ 1+\alpha} \left[2\dot H+3H^2 (1+\alpha )+\frac{1}{a^2} (1+3\alpha)\right], \hspace{1cm}
\end{equation} 
\begin{equation}\label{m6eq5_13}
 \rho=\frac{1}{ 8\pi G}\left(3H^2+\frac{3}{a^2}-\Lambda\right),\hspace{1cm}p=\alpha \rho.
\end{equation} 
Therefore, using Eqs.(\ref{m6eq3.35}) in Eq.(\ref{m6eq10_2}), we obtain the cosmological "constant"  
\begin{equation}\label{m6eq3.25_1}
\Lambda(t)=\frac{3}{4}t^2\left(2A-t\right)^2+ \frac{2(A-t) }{1+\alpha}
+ \frac{1}{a_0^2} \left( \frac{1+3\alpha}{1+\alpha}\right)  \exp \left[ \frac{1}{3}t^2(t-3A)\right]. 
\end{equation}
The value of the cosmological constant at the maximum life-time is
\begin{equation}\label{m6eq3.25}
\Lambda(2T)=48A^4-\frac{6A}{1+\alpha}+ \frac{1}{a_0^2}\left ( \frac{1+3\alpha}{1+\alpha}\right) \exp \left[ \frac{16}{3}A^3\right]. 
\end{equation}
And, using Eqs.(\ref{m6eq3.35}) in Eq.(\ref{m6eq5_13}), we obtain the density and pressure
\begin{subequations}
\begin{align}
\rho(t)&=\frac{1}{4\pi G(1+\alpha)}\left\{(t-A)+\frac{1}{a_0^2} \exp\left[ \frac{1}{3} t^2(t-3A)\right]\right\},\\
p(t)&=\alpha \rho(t).
\end{align}
\end{subequations}
The value of the pressure at the maximum life-time is
\begin{equation}
p(2T)=\frac{\alpha}{4\pi G(1+\alpha)}\left\{3A+\frac{1}{a_0^2} \exp\left[ \frac{16}{3} A^3 \right]\right\}.
\end{equation}
\section{Conclusion}
A model of a closed universe with a time-varying $\Lambda$ is considered. We constructed an optimal control problem for a closed universe with maximum life-time. The solution has been obtained by using Pontryagin's maximum principle. This approach may be used for obtaining new cosmological models with other interesting criteria.

\end{document}